\documentclass[12pt,a4paper]{article}
\usepackage{graphicx}
\usepackage{color}
\usepackage{bm}
\usepackage{amssymb,amsfonts,amsmath} 
\usepackage{cite}
\newcounter{figcount}

\begin{document}

\newcommand{\EQ}{Eq.~}
\newcommand{\EQS}{Eqs.~}
\newcommand{\FIG}{Fig.~}
\newcommand{\FIGS}{Figs.~}

\title{A network-based dynamical ranking system for competitive sports}
\author{Shun Motegi${}^{1}$ and Naoki Masuda${}^{1,2*}$
\\
\ \\
\ \\
${}^{1}$ 
Department of Mathematical Informatics,\\
The University of Tokyo,\\
7-3-1 Hongo, Bunkyo, Tokyo 113-8656, Japan
\ \\
${}^2$
PRESTO, Japan Science and Technology Agency,\\
4-1-8 Honcho, Kawaguchi, Saitama 332-0012, Japan\\
\ \\
$^*$ Corresponding author (masuda@mist.i.u-tokyo.ac.jp)}
\maketitle

\section*{Abstract}
From the viewpoint of networks, a ranking system for players or teams in sports is equivalent to a centrality measure for sports networks, whereby a directed link represents the result of a single game. Previously proposed network-based ranking systems are derived from static networks, i.e., aggregation of the results of games over time. However, the score of a player (or team) fluctuates over time. Defeating a renowned player in the peak performance is intuitively more rewarding than defeating the same player in other periods. To account for this factor, we propose a dynamic variant of such a network-based ranking system and apply it to professional men's tennis data. We derive a set of linear online update equations for the score of each player. The proposed ranking system predicts the outcome of the future games with a higher accuracy than the static counterparts.

\newpage

\section{Introduction}\label{sec:introduction}

Ranking of individual players or teams in sports, both professional
and amateur, is a tool for entertaining fans and developing sports
business.  Depending on the type of sports, different ranking systems
are in use \cite{Stefani1997JAS}.  A challenge in sports ranking is
that it is often impossible for all the pairs of players or teams (we
refer only to players in the following.  However, the discussion also
applies to team sports) to fight against each other. This is the case
for most individual sports and some team sports in which a league
contains many teams, such as American college football
and soccer at an international level.  Then, the
set of opponents depends on players such that ranking players by
simply counting the number of wins and losses is inappropriate.
 
In this situation, several ranking systems on the basis of
networks have been proposed. A player is regarded to be a node in
a network, and a directed link from the winning player
to the losing player (or the converse) represents the result of a single
game. Once the directed network of players is generated, 
ranking the players is equivalent to defining a
centrality measure for the network. 
A crux in constructing a network-based ranking system is to let
a player that beats a strong player gain a high score.
Examples of network-based ranking systems include
those derived from the Laplacian matrix of the network
\cite{Daniels1969Biom,Moon1970SiamRev,Borm2002AOR,Saavedra2010PhysicaA}, the PageRank
\cite{Radicchi2011PlosOne}, a random walk that is different from those
implied by the Laplacian or PageRank \cite{Callaghan2004NAMS},
a combination of node degree and global structure of networks
\cite{Herings2005SCW}, and the so-called win-lose score
\cite{Park2005JSM}.

Previous network-based ranking systems do not account for
fluctuations of rankings.
In fact, a player, even a history making strong player, referred to as
$X$, is often weak
in the beginning of the career. Player $X$ may also be
weak past the most brilliant period in the $X$'s career,
suggestive of the retirement in a near future.
For other players, it is more rewarding to beat $X$
when $X$ is in the peak performance than when $X$ is novice,
near the retirement, or in the slump.
It may be preferable to take into account the
dynamics of players' strengths for defining a ranking system.
In the present study, we extend the win-lose score, a network-based ranking system proposed by Park and Newman \cite{Park2005JSM} to the dynamical case.
Then, we apply the proposed ranking system
to the professional men's tennis data.

In broader contexts, the current study is related to at least two other
lineages of researches.
First, a dynamic network-based ranking implies that we exploit the
temporal information about the data, i.e., the times when games are played.
Therefore, such a ranking system is equivalent to
a dynamic centrality measure for temporal networks, in which 
sequences of pairwise interaction events with time stamps are building units
of the network
\cite{HolmeSaramaki2012PhysRep}.
Although some centrality measures specialized in temporal
networks have been proposed \cite{Tang2010SNS,Pan2011PRE,Grindrod2011PRE},
they are not for ranking purposes.
In addition, they are constant valued centrality measures for dynamic (i.e., temporal) data of pairwise interaction.
In the context of temporal networks, we propose
a dynamically changing centrality measure 
for temporal networks.

Second, statistical approaches to sports ranking have a much longer history
than network
approaches. Representative statistical ranking systems include the Elo
system \cite{Elo1978book} and the Bradley-Terry model (see
\cite{Bradley1976Biom} for a review).  Variants of these models have
been used to construct dynamic ranking systems.  Empirical Bayes
framework naturally fits this problem
\cite{Glickman1993thesis,Fahrmeir1994JASA,Glickman1999JRSSSC,Knorrheld2000Stat,Coulom2008LNCS,Herbrich2007NIPS}. Because the Bayesian estimators cannot be obtained analytically, or
even numerically owing to the computational cost, in these models,
techniques for obtaining Bayes estimators such as the Gaussian
assumption of the posterior distribution
\cite{Glickman1999JRSSSC,Herbrich2007NIPS}, approximate message
passing \cite{Herbrich2007NIPS}, and Kalman filter
\cite{Fahrmeir1994JASA,Glickman1999JRSSSC,Knorrheld2000Stat}, have
been employed. In a non-Bayesian statistical ranking system,
the pseudo likelihood, which is defined such that the contribution of the
past game results to the current pseudo likelihood decays
exponentially in time,
is numerically maximized \cite{Dixon1997ApplStat}.

In general, the parameter set of a statistical ranking system that accounts
for dynamics of players' strengths is composed of dynamically
changing strength parameters for all the players and perhaps other
auxiliary parameters. 
Therefore, the number of parameters to be statistically estimated
may be large relative to the amount of data.
In other words, the instantaneous ranks of players have to be
estimated before the players play sufficiently many games with others under
fixed strengths.
Even under a Bayesian framework with which updating of the parameter values
is naturally implemented, it may be difficult to reliably estimate
dynamic ranks of players due to relative paucity of data.
In addition, in sports played by individuals, such as tennis, 
it frequently occurs that new players begin and old and underperforming 
players leave. This factor also increases the
number of parameters of a ranking system.
In contrast, ours and other
network-based ranking systems, both static and dynamic ones,
are not founded on statistical methods.
Network-based ranking systems
can be also simpler and more transparent than statistical counterparts.

\section*{Results}\label{sec:results}

\subsection*{Dynamic win-lose score}\label{sub:dwl}

We extend the
win-lose score \cite{Park2005JSM} (see Methods) to
account for the fact that the strengths of players fluctuate over time.
In the following, we refer to the win-lose score as the original
win-lose score and the extended one as the dynamic
win-lose score.

The original win-lose score overestimates
the real strength of a player $i$ when $i$ defeated an opponent $j$ that is
now strong and was weak at the time of the match between $i$ and $j$.
Because $j$ defeats many strong opponents afterward,
$i$ unjustly receives many indirect wins through $j$.
The same logic also applies to other network-based
static ranking systems \cite{Daniels1969Biom,Moon1970SiamRev,Borm2002AOR,Saavedra2010PhysicaA,Radicchi2011PlosOne,Callaghan2004NAMS,Herings2005SCW}.

To remedy this feature, we pose two assumptions.
First, we assume that
the increment of the win score of player $i$ through the $i$'s winning
against player $j$ depends on the $j$'s win score at that moment. It
does not explicitly depend on the $j$'s score in the past or future.
The same holds true for the lose score.
Second, we assume that each player's win and lose scores
decay exponentially in time. This assumption is also employed in a
Bayesian dynamic ranking system \cite{Dixon1997ApplStat}.

Let $A_{t_n}$ be the win-lose matrix for the game that occurs at time $t_n$
($1\le n\le n_{\max}$).
In the analysis of the tennis data carried out in the following,
the resolution of $t_n$ is equal to one day. Therefore, players'
scores change even within a single tournament.
If player $j$ wins against player $i$ at time $t_n$,
we set the $(i,j)$ element of the matrix $A_{t_n}$ to be 1. All the other elements of $A_{t_n}$
are set to 0.
We define the dynamic win score at time $t_n$ in vector form, denoted
by $\bm w_{t_n}$, as follows:
\begin{align}
W_{t_n} =& A_{t_n} + e^{-\beta (t_n - t_{n-1})} \sum _ {m_n \in \{ 0,1 \}} \alpha^{m_n} A_{t_{n-1}}A_{t_n}^{m_n} \notag\\ 
&+ e^{-\beta (t_n - t_{n-2})} \sum _ {m_{n-1},m_n \in \{ 0,1 \}} \alpha^{m_{n-1}+m_n} A_{t_{n-2}}A_{t_{n-1}}^{m_{n-1}}A_{t_n}^{m_n}  \notag\\ 
&+ \cdots + e^{-\beta (t_n - t_1)} \sum _ {m_2, \ldots, m_n \in \{ 0,1 \}}
\alpha^{\sum_{i=2}^n m_i} A_{t_1}A_{t_2}^{m_2}\cdots A_{t_n}^{m_n}
\label{eq:def of W_{t_n}}
\end{align}
and
\begin{equation}
\bm w_{t_n} = W_{t_n}^{\top}\bm 1,
\label{eq:def of w_{t_n}}
\end{equation}
where $\alpha$ is the weight of the indirect win, which is the same as
the case of the original win-lose score (Methods),
and $\beta\ge 0$ represents the decay rate of the score.

The first term on the right-hand side of Eq.~\eqref{eq:def of
  W_{t_n}} (i.e., $A_{t_n}$) represents the effect of the direct win at 
time $t_n$.  The second term consists of two contributions. For
$m_n=0$, the quantity inside the summation represents the direct win
at time $t_{n-1}$, which results in weight $e^{-\beta (t_n-t_{n-1})}$.
For $m_n=1$, the quantity represents the
indirect win. The ($i$, $j$) element of $A_{t_{n-1}}A_{t_n}$ is positive
if and only if 
player $j$ wins against a player $k$ at time $t_n$ and $k$ wins 
against $i$ at time $t_{n-1}$. Player $i$ gains score 
$e^{-\beta (t_n-t_{n-1})} \alpha$ out of this situation.
For both cases $m_n=0$ and $m_n=1$,
the $j$th column of the second term accounts for the effect of the
$j$'s win at time $t_{n-1}$.
The third term covers four cases. For $m_{n-1}=m_n=0$,
the quantity inside the summation represents the direct win at $t_{n-2}$,
resulting in weight $e^{-\beta (t_n-t_{n-2})}$. For $m_{n-1}=0$ and $m_n=1$, 
the quantity represents the indirect win based on the games at $t_{n-2}$ and
$t_n$, resulting in weight $e^{-\beta (t_n-t_{n-2})}\alpha$.
For $m_{n-1}=1$ and $m_n=0$, the quantity represents the indirect win
based on the games at $t_{n-2}$ and $t_{n-1}$, resulting in weight 
$e^{-\beta (t_n-t_{n-2})}\alpha$. For $m_{n-1}=m_n=1$, the quantity
represents the indirect win based on the games at $t_{n-2}$,
$t_{n-1}$, and $t_n$, resulting in weight
$e^{-\beta (t_n-t_{n-2})}\alpha^2$.
In either of the four cases, the $j$th column of the third term accounts for the effect of the $j$'s win at time $t_{n-2}$.

\begin{figure}
\begin{center}
\includegraphics[width=12cm]{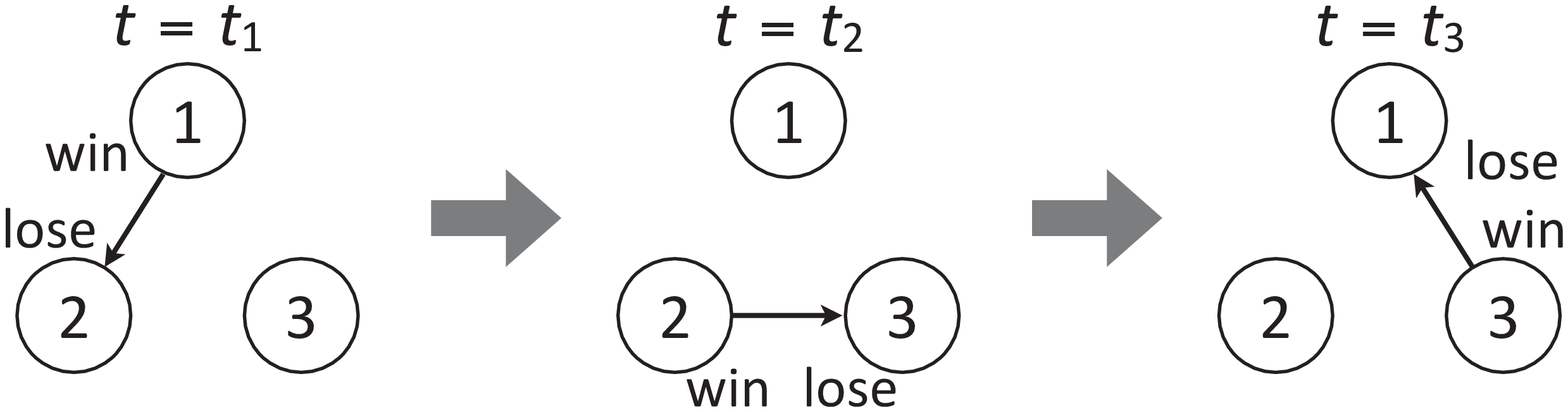}
\caption{Example time series of games with $N=3$.}
\label{fig:N=3 example}
\end{center}
\end{figure}

To see the difference between the original and dynamic win scores,
consider the exemplary data with $N=3$ players shown in
\FIG\ref{fig:N=3 example}. The original
win-lose scores calculated from the aggregation of the data up 
to time $t_n$ ($n=1, 2$, and 3), denoted by $w_{t_n}(i)$ for player $i$, are given by
\begin{equation}
\begin{cases}
w_{t_1}(1)=1,\\
w_{t_1}(2)=0,\\
w_{t_1}(3)=0,
\end{cases}
\begin{cases}
w_{t_2}(1)=1+\alpha,\\
w_{t_2}(2)=1,\\
w_{t_2}(3)=0,
\end{cases}
\begin{cases}
w_{t_3}(1)=1+\alpha+\alpha^2+\cdots,\\
w_{t_3}(2)=1+\alpha+\alpha^2+\cdots,\\
w_{t_3}(3)=1+\alpha+\alpha^2+\cdots.
\end{cases}
\end{equation}
The scores of the three players are the same at $t=t_3$ because the aggregated network
is symmetric (i.e., directed cycle) if we discard the information
about the time.

The dynamic win-lose scores for the same data are given by
\begin{equation}
\begin{cases}
w_{t_1}(1)=1,\\
w_{t_1}(2)=0,\\
w_{t_1}(3)=0,
\end{cases}
\begin{cases}
w_{t_2}(1)=e^{-\beta(t_2-t_1)},\\
w_{t_2}(2)=1,\\
w_{t_2}(3)=0,
\end{cases}
\begin{cases}
w_{t_3}(1)=e^{-\beta(t_3-t_1)},\\
w_{t_3}(2)=e^{-\beta(t_3-t_2)},\\
w_{t_3}(3)=1+\alpha e^{-\beta(t_3-t_1)}.
\end{cases}
\end{equation}
The score of player 1 at $t_2$ (i.e., $w_{t_2}(1)$) 
differs from the original win-lose score in two aspects.
First, it is discounted by factor
$e^{-\beta(t_2-t_1)}$. Second, the value of $w_{t_2}(1)$ indicates that
player 1 does not gain an indirect win. This is
because it is after player 1 defeated player 2 that player 2
defeats player 3.
In contrast, player 3 gains an indirect win at $t=t_3$ because player 3
defeats player 1, which defeated player 2 before (i.e.,
at $t=t_1$).
It should be noted that the win scores of the three players are
different at $t=t_3$ although the aggregated network is symmetric.

Equation~\eqref{eq:def of W_{t_n}} leads to
\begin{align}
W_{t_n} =& A_{t_n} + e^{-\beta (t_n - t_{n-1})} \left[ A_{t_{n-1}} 
+ e^{-\beta (t_{n-1} - t_{n-2})} \sum_{m_{n-1} \in \{ 0,1 \}} 
\alpha^{m_{n-1}} A_{t_{n-2}}A_{t_{n-1}}^{m_{n-1}}\right.\notag\\ 
&+\cdots\notag\\
& \left. + e^{-\beta (t_{n-1} - t_1)} \sum_{m_2, \ldots, m_{n-1} \in \{ 0,1
  \}} \alpha^{\sum _{i=2}^{n-1} m_i} A_{t_1}A_{t_2}^{m_2}\cdots A_{t_{n-1}}^{m_{n-1}} \right] \sum_{m_n \in \{ 0,1 \}} \alpha^{m_n} A_{t_n}^{m_n}\notag\\
=& A_{t_n} +  e^{-\beta (t_n - t_{n-1})}W_{t_{n-1}}(I + \alpha A_{t_n}).
\label{eq:convenient W_{t_n}}
\end{align}
Therefore, by combining Eqs.~\eqref{eq:def of w_{t_n}} and \eqref{eq:convenient W_{t_n}}, we obtain
the update equation for the dynamic win score as follows:
\begin{equation}
\bm w_{t_n}=\begin{cases}
A_{t_1}^{\top}\bm 1 & (n=1),\\
A_{t_n}^{\top} \bm 1 + e^{-\beta (t_n - t_{n-1})} (I + \alpha
A_{t_n}^{\top})\bm w_{t_{n-1}} & (n>1).
\end{cases}
\label{eq:wtn}
\end{equation}
The dynamic lose score at time $t_n$ is denoted
in vector form by $\bm \ell_{t_n}$.
We obtain the update equation for $\bm \ell_{t_n}$
by replacing $A_{t_n}$ in \EQ\eqref{eq:wtn} by $A_{t_n}^{\top}$
as follows:
\begin{equation}
\bm \ell_{t_n}=\begin{cases}
A_{t_1}\bm 1 & (n=1), \\
A_{t_n} \bm 1 + e^{-\beta (t_n - t_{n-1})} (I + \alpha A_{t_n})\bm \ell_{t_{n-1}} & (n>1).\end{cases}
\label{eq:ltn}
\end{equation}
Finally, the dynamic win-lose score at time $t_n$, denoted by $\bm s_{t_n}$, is
given by
\begin{equation}
\bm s_{t_n} = \bm w_{t_n} -\bm \ell_{t_n}.
\end{equation}

It should be noted that we do not treat retired players in special ways. Players' scores exponentially decay after retirement.

\subsection*{Predictability}\label{sub:predictability}

We apply the dynamic win-lose score to results of professional men's tennis.
The nature of the data is described in Methods.

In this section, we predict the outcomes of future games
based on different ranking systems.
The frequency of violations, whereby
a lower ranked player wins against a higher ranked
player in a game, quantifies the degree of predictability
\cite{Martinich2002Inter,Bennaim2006JQAS}.
In other literature, 
the retrodictive version of the frequency of violations is also used
for assessing the performance of ranking systems
\cite{Martinich2002Inter,Lundh2006JQAS,Park2010JSM,Coleman2005Interface}.

We compare the predictability of the 
dynamic win-lose score, the original win-lose score
\cite{Park2005JSM}, and the prestige score (Methods).
The prestige score, proposed by Radicchi
and applied to
professional men's tennis data \cite{Radicchi2011PlosOne},
is a static ranking system and
is a version of the PageRank originally proposed for ranking webpages
\cite{Brin98}. We also implement a dynamic version of the prestige score (Methods) and compare its performance of prediction with that of the dynamic win-lose score.

We define the frequency of violations as follows.
We calculate the score of each player at $t_n$ ($1\le n\le n_{\max}-1$)
on the basis of the results up to $t_n$.
For the original win-lose score and prestige score, we
aggregate the directed links
from $t=t_1$ to $t=t_n$ to construct a static network and
calculate the players' scores.
If the result of each game at $t_{n+1}$ is
inconsistent with the calculated ranking,
we regard that a violation occurs.
If the two players involved in the game at $t_{n+1}$
have exactly the same score,
we regard that a tie occurs irrespective of the result of the game.
We define the prediction accuracy at the $N_{\rm gp}$th game
as the fraction of correct prediction when the results of the
games from $t=t_2$ through the $N_{\rm gp}$th game
are predicted. The prediction accuracy is given by
$\left(N_{\rm gp}^{\prime}-e-v\right)/\left(N_{\rm gp}^{\prime}-e\right)$,
where $N_{\rm gp}^{\prime} (<N_{\rm gp})$ is the number of predicted games,
$v$ is the number of violations, and $e$ is the number of ties.

For the prestige score and its dynamic variant,
 we exclude the games in which either
player plays for the first time because the score is not defined for
the players that have never played. In this case, we increment $e$ by
one.  

The original and dynamic win-lose scores can be negative
valued. Equations~\eqref{eq:def of w_{t_n}}
and \eqref{eq:w original win score} guarantee that the initial score is equal to zero for all the players for the dynamic and original win-lose scores, respectively.
Furthermore, any player has a zero win-lose score
when the player fights a game for the first time.
Even though we do not treat such a game
as tie unless both players involved in the game have zero scores, treating it as tie
little affects the following results.

\begin{figure}
\begin{center}
\includegraphics[width=16cm]{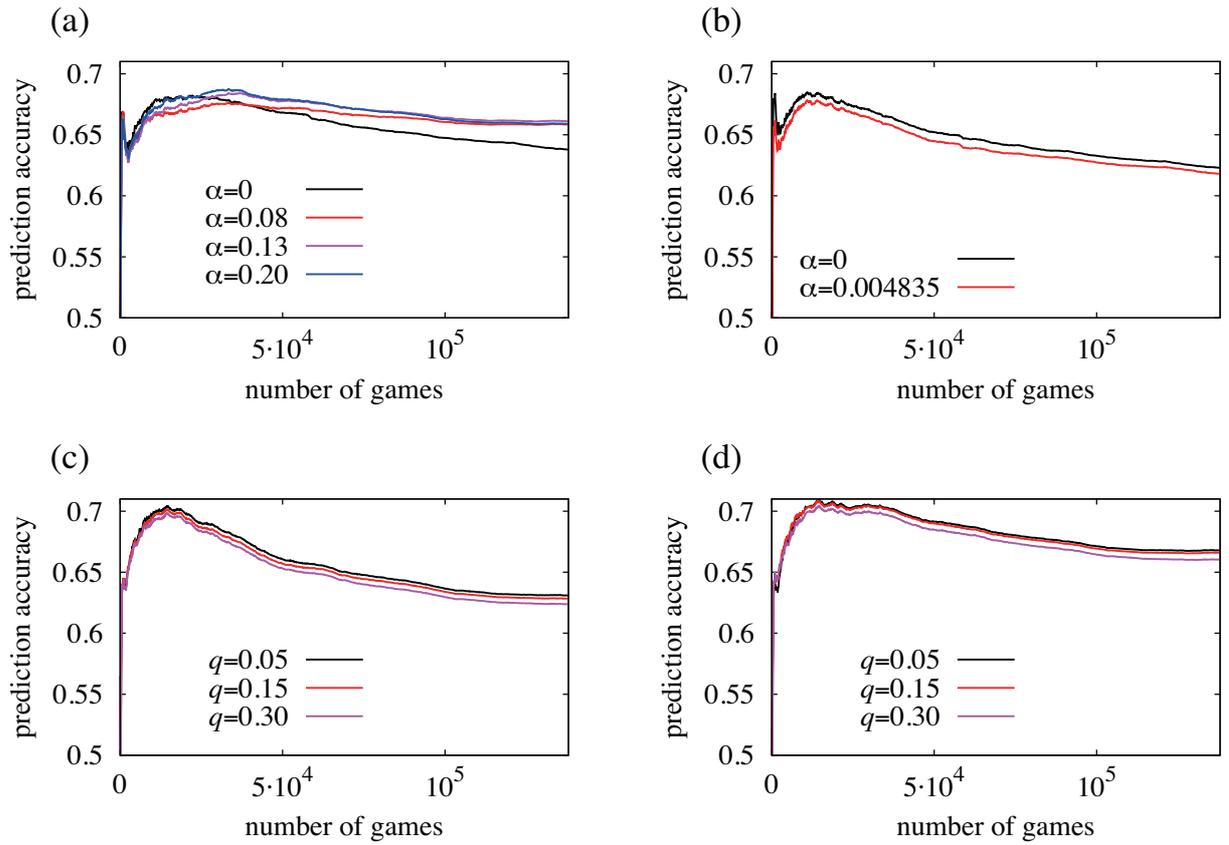}
\caption{Performance of prediction for the four ranking systems. (a) Dynamic win-lose score with $\beta=1/365$ and different $\alpha$ values. (b) Original win-lose score with $\alpha=0$ and 0.004835. (c) Prestige score with $q=0.05$, 0.15, and 0.3. (d) Dynamic variant of the prestige score with $q=0.05$, 0.15, and 0.3.}
\label{fig:performance}
\end{center}
\end{figure}

The prediction accuracy for the dynamic win-lose score,
original win-lose score, prestige score, and dynamic prestige score
are shown in 
\FIGS\ref{fig:performance}(a), \ref{fig:performance}(b),
\ref{fig:performance}(c),
and \ref{fig:performance}(d), respectively, for various parameter values.

Figure~\ref{fig:performance}(a) indicates that the prediction accuracy
for the dynamic win-lose score is the largest for
$\alpha=0.13$ except when the number of games (i.e., $N_{\rm gp}$) is
small. The accuracy is insensitive to $\alpha$ when $0.08\le \alpha\le
0.2$. In this range of $\alpha$, we confirmed by additional numerical simulations that
the results for $\beta=1/365$ and
those for $\beta=0$ are indistinguishable. Therefore, we
conclude that the performance of prediction has some robustness with
respect to $\alpha$ and $\beta$.  We also confirmed that
the accuracy monotonically increases between $\alpha\approx 0.03$ and
$\alpha\approx 0.13$.  However, for an unknown reason, the accuracy
with $\alpha\approx 0.03$ is smaller than that with $\alpha=0$ (results
not shown).

Figure~\ref{fig:performance}(b) indicates that the prediction accuracy
for the original win-lose score is larger for $\alpha=0$
than $\alpha=0.004835$. The latter
$\alpha$ value is very close to the upper
limit calculated from the largest eigenvalue of $A$ (see subsection ``Parameter values'' in Methods). We also found that the prediction accuracy monotonically 
decreases with
$\alpha$. Nevertheless, except for small $N_{\rm gp}$,
the accuracy with
 $\alpha=0$ is lower than that for the dynamic win-lose score
with $\alpha=0$ and $0.08\le \alpha\le 0.2$ (\FIG\ref{fig:performance}(a)).

Figure~\ref{fig:performance}(c) indicates that
the prediction by the prestige score is better for a smaller
value of $q$ (see Methods for the meaning of $q$).
We confirmed that this is the case for other values of
$q$ and that the results with $q\le 0.05$ little
differ from those with $q=0.05$. 
Except for small $N_{\rm gp}$, the prediction
accuracy with $q=0.05$ is lower than that for the dynamic win-lose score with
$0.08\le \alpha\le 0.2$ (\FIG\ref{fig:performance}(a)).

Figure~\ref{fig:performance}(d) indicates that
the prediction by the dynamic variant of the prestige score is more accurate than that by the dynamic win-lose score, in particular for small $N_{\rm gp}$. Similar to the case of the original prestige score, the prediction accuracy decreases with $q$.

The findings obtained from \FIG\ref{fig:performance} are summarized as follows. When
$\alpha$ is between $\approx 0.08$ and $\approx 0.2$ and 
$\beta$ is between 0 and $1/365$,
the dynamic win-lose score outperforms the original win-lose score and the prestige score in the prediction accuracy. For example,
at the end of the data,
the accuracy is equal to 0.659, 0.661, 0.661, and 0.659
for the dynamic win-lose score with
($\alpha$, $\beta$) $=$ (0.08, $1/365$),
 (0.1, $1/365$), (0.13, $1/365$), and (0.2, $1/365$), respectively, 
while it is equal to 0.623 for the original win-lose score with $\alpha=0$
and 0.631 for the prestige score with $q=0.05$. 
However, the accuracy for
the dynamic variant of the prestige score with $q=0.05$ (i.e., 0.668) is slightly larger than the largest value obtained by the dynamic win-lose score.

We also compare the prediction accuracy for the dynamic win-lose score with that for the official Association of Tennis
Professionals (ATP) rankings. Because the calculation of the ATP rankings involves relatively minor games that do not belong to ATP World Tour tournaments, which we used for \FIG\ref{fig:performance}, we use a different data set for the present comparison (see ``Data'' in Methods). The prediction accuracy at the end of the data is equal to 0.637 for the ATP rankings and 0.588, 0.629, 0.646, 0.650, and 0.649 for the dynamic win-lose score with
($\alpha$, $\beta$) $=$ (0.08, $1/365$),
 (0.1, $1/365$), (0.13, $1/365$), (0.17, $1/365$), and (0.2, $1/365$), respectively. The prediction accuracy for the dynamic win-lose score is larger than that for the ATP rankings in a wide range of $\alpha$ (i.e., $0.11\le \alpha\le 0.39$).

\subsection*{Robustness against parameter variation}\label{sub:sensitivity}

Figure~\ref{fig:performance}(a) indicates that 
the prediction accuracy for the dynamic win-lose score is robust against some variations in the $\alpha$ and $\beta$ values.
In this section, we examine the robustness of the dynamic win-lose score
more extensively by examining the rank correlation between
the scores derived from different $\alpha$ and $\beta$ values.

The Kendall's tau is a standard method to quantify the rank
correlation \cite{Kendall1938Biom}. In our data,
the full ranking containing all the players, to which the Kendall's
tau applies, contains players that only appear in a few games.
In fact, most players are such players \cite{Radicchi2011PlosOne},
and their ranks are inherently
unstable. In addition, it is usually the list of top ranked players that are
of practical interests.

Therefore, 
we use a generalized Kendall's tau for comparing top $k$ lists
of the full ranking \cite{Fagin2003SIAMDM}.
We denote the sets of the top $k$ players, i.e., $k$ players with the
largest scores, in the two full rankings by
$\bm R_1$ and $\bm R_2$. 
In general, $\bm R_1$ and $\bm R_2$ can be different.
For an arbitrarily chosen pair of players
$r_1$, $r_2$ $\in \bm R_1 \cup \bm R_2$, $r_1\neq r_2$,
we set $\overline{K}_{r_1 , r_2}(\bm R_1,\bm R_2)=1$ if
(1) $r_1$ and $r_2$ appear in both top $k$ lists $R_1$ and $R_2$, and $r_1$ and $r_2$ are in the opposite order in the two top $k$ lists,
(2) $r_1$ has a higher rank than $r_2$ in one of the top $k$ lists, and
$r_2$, but not $r_1$, is contained in the other top $k$ list,
(3) $r_1$ exists only in one of the two top $k$ lists, and $r_2$
exists only in the other top $k$ list.
Otherwise, we set $\overline{K}_{r_1, r_2}(\bm R_1,\bm R_2)=0$.
$\overline{K}_{r_1, r_2}(\bm R_1,\bm R_2)$ is a penalty imposed on the inconsistency between the two top $k$ lists. We use the so-called
optimistic variant of the Kendall distance
$K_{\tau}^{(0)}(\bm R_1,\bm R_2)$ defined as follows \cite{Fagin2003SIAMDM}:
\begin{equation}
K_{\tau}^{(0)}(\bm R_1,\bm R_2) = \sum_{r_1,r_2 \in \bm R_1 \cup \bm R_2} \overline{K} _{r_1, r_2}(\bm R_1,\bm R_2).
\end{equation}
We normalize the distance between the two rankings as follows
\cite{Mccown2007JCDL}:
\begin{equation}
K = 1 - \frac{K_{\tau}^{(0)}(\bm R_1,\bm R_2)}{k^2}.
\end{equation}
A large value of $K$ indicates a higher correlation between the two
top $k$ lists.
It should be noted that $0\le K\le 1$. In particular, when there is no overlap
between the two top $k$ lists,
we obtain $K=0$.

\begin{figure}
\begin{center}
\includegraphics[width=12cm]{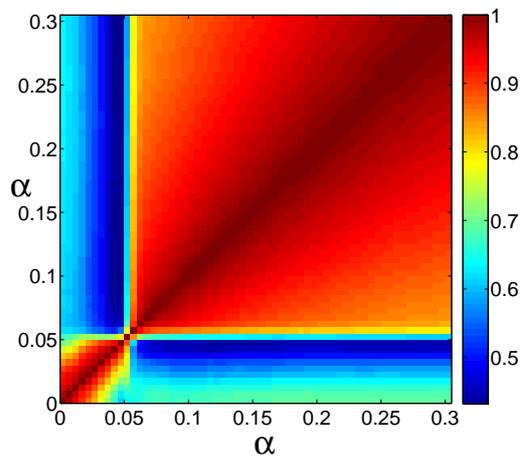}
\caption{Rank correlation between the two top 300 lists for the dynamic win-lose score with $\beta=1/365$. Pairs of rankings with different values of $\alpha$ are compared.}
\label{fig:robustness alpha}
\end{center}
\end{figure}

For the dynamic win-lose scores at $t_{n_{\max}}$,
i.e., at the end of the entire period,
we calculate $K$ with $k=300$ for different pairs of $\alpha$ and $\beta$
values.
The results for $\beta=1/365$ and 
different values of $\alpha$ are shown in
\FIG\ref{fig:robustness alpha}. The top $k$ lists are similar (i.e., $K \ge 0.85$) for any $\alpha$ larger than $\approx 0.06$.
This finding is consistent with the fact
that the prediction accuracy is high
and robust when $\alpha$ falls between $\approx 0.08$ and $\approx 0.2$
(\FIG\ref{fig:performance}(a)).

\begin{figure}
\begin{center}
\includegraphics[width=12cm]{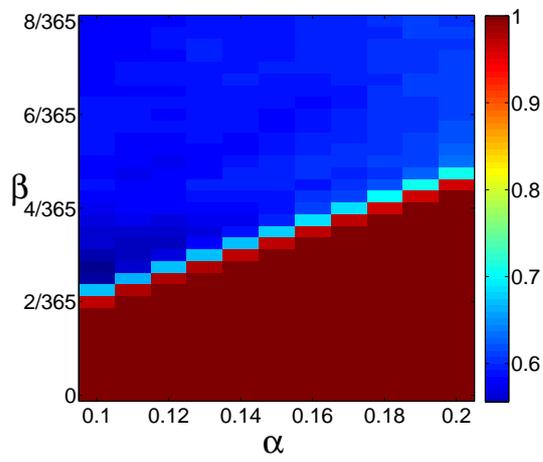}
\caption{Rank correlation between the two top 300 lists for the dynamic win-lose score with fixed $\alpha$ values. Pairs of ranking, one with $\beta=1/365$ and the other with a general $\beta$ value, are compared.}
\label{fig:robustness alpha beta}
\end{center}
\end{figure}

For fixed values of $\alpha$, the $K$ values between the ranking
with $\beta=1/365$ and that with various values of $\beta$ are shown
in \FIG\ref{fig:robustness alpha beta}. $K$ is almost
unity at least in the range $0\le\beta\le 2/365$.  Therefore, removing
the assumption of the exponential
decay of score in time (i.e., $\beta=0$) little changes the top
300 list.  This finding is consistent with the result
that the prediction accuracy is almost the same between $\beta=0$ and
$\beta=1/365$ if $0.1\le\alpha\le 0.2$ (see the previous subsection).
Nevertheless, this observation does not imply that we can ignore the
temporal aspect of the data. Keeping the order of the
games contributes to the performance of prediction, as suggested by
the comparison between the prediction results for
the dynamic (\FIG\ref{fig:performance}(a))
and original (\FIG\ref{fig:performance}(b)) win-lose scores.

\subsection*{Dynamics of scores for individual players}\label{sub:dynamics individual}

\begin{figure}
\begin{center}
\includegraphics[width=12cm]{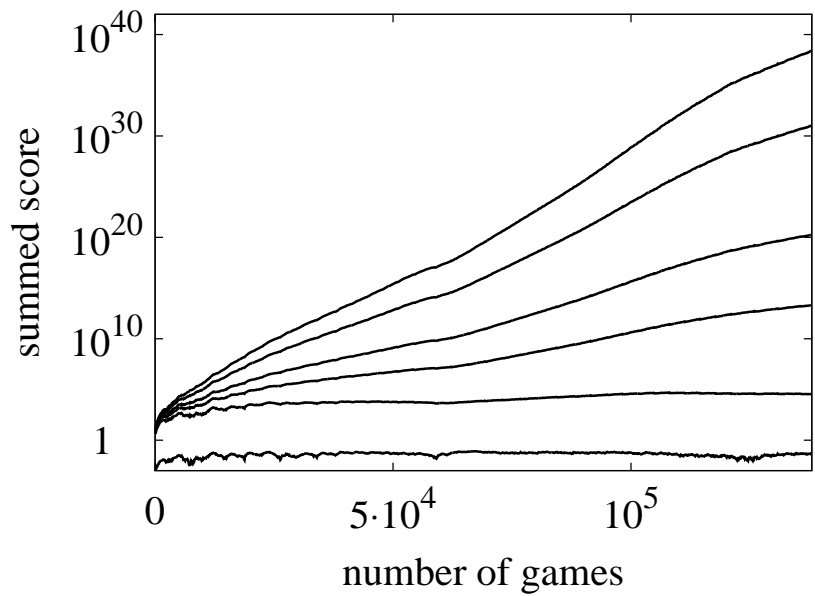}
\caption{Dynamics of the summation of the scores (i.e., $\sum_{i=1}^N s_{t_n}(i)$). The lines correspond to $\alpha=0.15$ (top), 0.13, 0.1, 0.08, 0.05, and $10^{-5}$ (bottom). The results for $\alpha=0$ are not shown because $\sum_{i=1}^N s_{t_n}(i)$ often takes negative values.}
\label{fig:sum scores}
\end{center}
\end{figure}

In contrast to the original win-lose score and prestige score, the
dynamic win-lose score can track dynamics of the strength of each
player. It should be noted that the summation of the scores over the
individuals, i.e., $\sum_{i=1}^N s_{t_n}(i)$, depends on time. In particular,
it grows almost exponentially for the parameter values with which the
prediction accuracy is high (i.e., $\alpha$ larger than $\approx
0.08$), as shown in \FIG\ref{fig:sum scores}.  $\sum_{i=1}^N s_{t_n}(i)$
increases with the number of games, or equivalently, with time because
more recent players take more advantage of indirect wins than older
players. 
The increase in $\sum_{i=1}^N s_{t_n}(i)$ is not owing to 
the number of players or games observed per year;
in fact, the latter numbers do not increase in time \cite{Radicchi2011PlosOne}.

\begin{figure}
\begin{center}
\includegraphics[width=16cm]{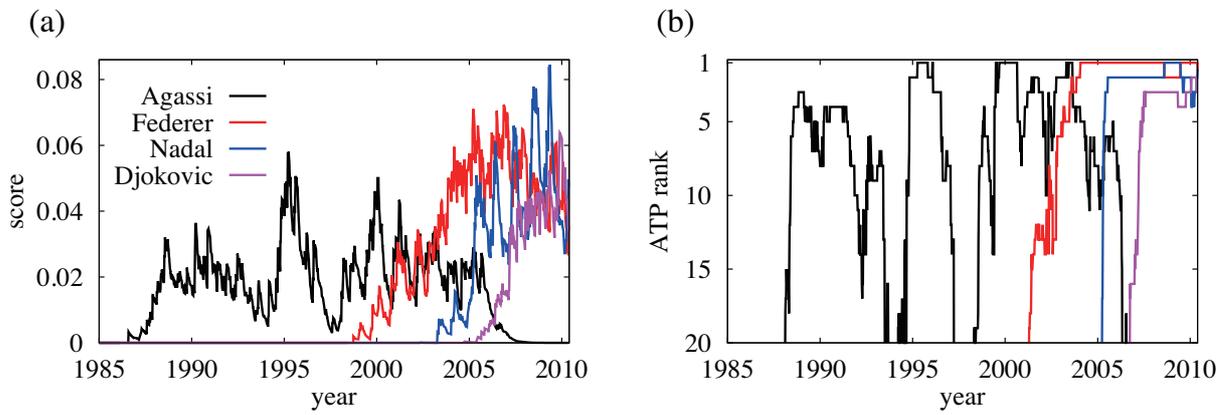}
\caption{(a) Time courses of normalized win-lose scores for Andre Agassi, Roger Federer, Rafael Nadal, and Novak Djokovic. (b) Time courses of the ATP rankings for the four players.}
\label{fig:famous players}
\end{center}
\end{figure}

Therefore, for clarity, we normalize the win-lose score of each player by dividing it by the instantaneous $\sum_{i=1}^N s_{t_n}(i)$ value.
The time courses of the normalized win-lose scores for four
renowned players are shown in
\FIG\ref{fig:famous players}(a). 
We set $\alpha=0.13$ and $\beta=1/365$, for
which the prediction is approximately the most accurate. The ATP rankings of the four players during the same period are shown in \FIG\ref{fig:famous players}(b) for comparison. The time courses of 
the dynamic win-lose score and those of the ATP rankings are similar. 
In particular, the times at which the strength of one player (e.g., Federer) begins to exceed another player (e.g., Agassi) are similar between
\FIGS\ref{fig:famous players}(a) and \ref{fig:famous players}(b).
Figure~\ref{fig:famous players} suggests
that the dynamic win-lose score appositely captures
rises and falls of these players.

\section*{Discussion}

We extended the win-lose score for static sports networks
\cite{Park2005JSM} to the case of dynamic networks. By assuming that
the score decays exponentially in time, we could derive closed
online update equations for the win and lose scores. The proposed dynamic win-lose
score realizes a higher prediction accuracy than the original win-lose
score and the prestige score.
It is straightforward to extend the dynamic win-lose score to incorporate
factors such as the importance of each tournament or game via modifications 
of the game matrix $A_{t_n}$.
We also confirmed the robustness of the
ranking against variation in the two parameter values in the model.
Finally, the dynamic win-lose score is capable of
tracking dynamics of players' strengths.

It seems that network-based ranking systems are easier to understand and
implement, and more scalable than those based on statistical methods.
The dynamic win-lose score share these desirable features with
static network-based ranking systems.

The applicability of the idea behind the dynamic win-lose score is not 
limited to the case of the win-lose score. In fact, we implemented a dynamic variant of the prestige score. It even yielded a larger prediction accuracy than the dynamic win-lose score did. This result implies that
the idea of network-based dynamic ranking systems may be a powerful approach
to assessing strengths of sports players and teams, which fluctuate over time. The dynamic win-lose score is better than our version of the dynamic prestige score in that only the former allows for a set of closed online update equations. Establishing similar update equations
for other network-based ranking systems
such as the prestige score and the Laplacian
centrality (see Introduction) is warranted for future work.
Prospective results obtained through
this line of researches may be also useful in 
systematically deriving dynamic centrality measures for
temporal networks in general.

\section*{Methods}

\subsection*{Park \& Newman's win-lose score}\label{sub:Park}

The win-lose score by Park and Newman \cite{Park2005JSM} is a
network-based static ranking system defined as follows.
We assume $N$ players and denote by $A_{ij}$ ($1\le i, j\le N$)
the number of times that player $j$ wins against player $i$ during the entire period.
We let $\alpha$ ($0\le \alpha<1$) be a constant representing the
weight of indirect wins.
For example,
if player $i$ wins against $j$ and $j$ wins against $k$,
$i$ gains score 1 from the direct win against $j$ and score $\alpha$
from the indirect win against $k$. Therefore, the 
$i$'s win score is equal to $1+\alpha$.
If $k$ wins against yet another player $\ell$,
the $i$'s win score
is altered to $1+\alpha+\alpha^2$.

The win scores of the players are given by
\begin{align}
W =& A + \alpha A^2 + \alpha^2 A^3 + \cdots\notag\\ 
=& A(I+\alpha A +  \alpha^2 A^2 + \alpha^3 A^3 + \cdots)\notag\\
=& A(I - \alpha A)^{-1},\\ 
\bm w =& W^{\top} \bm 1 = (I-\alpha A^{\top})^{-1}A^{\top} \bm 1,
\label{eq:w original win score}
\end{align}
where $W$ is the $N\times N$ matrix whose $(i,j)$ element represents
the score that player $j$ obtains via
direct and indirect wins against player $i$,
$\bm w$ is the $N$ dimensional column vector whose $i$th
element represents the win score of player $i$, and
$\bm 1$ is the $N$ dimensional column vector defined by
\begin{equation}
\bm 1=(1\; 1\; \cdots \; 1)^{\top}. 
\end{equation}
We similarly obtain
the lose scores of the $N$ players
in vector form by replacing $A$ with $A^{\top}$ as follows:
\begin{equation}
\bm \ell = (I-\alpha A)^{-1}A \bm 1.
\end{equation}
The total win-lose score is given in vector form by
\begin{equation}
\bm s = \bm w -\bm \ell.
\end{equation}

\subsection*{Prestige score}

The prestige score of player $i$, denoted by $P_i$, is defined by
\begin{equation}
P_i = (1-q)\sum_{j=1}^N P_j\frac{\tilde{w}_{ji}}{s_j^{\rm out}} + \frac{q}{N} + \frac{1-q}{N}\sum_{j=1}^N P_j\delta (s_j^{\rm out})\quad (1\le i\le N),
\label{eq:prestige score}
\end{equation}
where $q$ is a constant,
$\tilde{w}_{ji}$ is the number of times player $i$ defeats player $j$ during the entire
period (it should be noted that $\tilde{w}_{ji}$ has nothing to do with the win scores denoted by $\bm w$ in \EQS\eqref{eq:def of w_{t_n}} and \eqref{eq:w original win score}),
$s_j^{\rm out}\equiv\sum_{i^{\prime}=1}^N \tilde{w}_{ji^{\prime}}$ is equal to the number of losses for player $j$, $\delta(s_j^{\rm out})=1$ if $s_j^{\rm out}=0$,
and $\delta(s_j^{\rm out})=0$ 
if $s_j^{\rm out}\ge 1$. The normalization is given by
$\sum_{i=1}^N P_i = 1$.
We set $q=0.15$, as in 
\cite{Radicchi2011PlosOne}, and also $q=0.05$ and $q=0.30$.

To define a dynamic variant of the prestige score,
we let $\tilde{w}_{ij}$ used in \EQ\eqref{eq:prestige score}
depend on time. We define $\tilde{w}_{ij}$
at time $t$ by
\begin{equation}
\tilde{w}_{ji}\equiv \sum_n A_{t_n}(j,i)e^{-\beta(t-t_n)},
\label{eq:tilde w(t)}
\end{equation}
where $A_{t_n}(j,i)$ is the $(j,i)$ element of the win-lose matrix $A_{t_n}$, and the summation over $n$ is taken over the games that occur before time $t$.
Substituting \EQ\eqref{eq:tilde w(t)} in
\EQ\eqref{eq:prestige score} yields the dynamic prestige score $P_i$ 
($1\le i\le N$) at time $t$. We set $\beta=1/365$, which is the same value as that used for the dynamic win-lose score.

\subsection*{Data}

We collected the data from the website of ATP \cite{ATPWorldTour}.
Except when we compared the prediction accuracy for the
dynamic win-lose score with that for the ATP rankings,
we used single games in
ATP World Tour tournaments recorded on this website.
The data set contains 137842 singles games from December 1972 to May
2010 and involves 5039 players that
participated in at least one game.
Because the source of our data set is the same as that of 
Radicchi's data set \cite{Radicchi2011PlosOne} and the 
period of the data is similar,
the number of games contained in our data and that in
Radicchi's are close to each other.

In the comparison between the dynamic win-lose score and the ATP rankings,
we used all the types of single games recorded on the website of ATP.
They include the games belonging to
ATP Challenger Tours and ITF Futures tournaments in addition to
ATP World Tour tournament games. We used this data set because it corresponds to the games on which the calculation of the ATP rankings is based. The ATP rankings are not available 
on a regular basis in early years. Therefore, we used the data from July 23, 1984 to August 15, 2011. The data set contains 330796 games and involves 13077 players
that participated in at least one game.

\subsection*{Parameter values for the dynamic win-lose score}\label{sub:parameter choice}

A guiding principle for setting
the parameter values of a ranking system is to select the values that 
maximize the performance of prediction
\cite{Dixon1997ApplStat,Knorrheld2000Stat}. Instead,
we set $\alpha$ and $\beta$ as follows.

In the original win-lose score, it is recommended that $\alpha$ is set
to the value smaller than and close to the inverse of the largest eigenvalue of
$A$ \cite{Park2005JSM}. If $\alpha$ exceeds this upper limit, the original win-lose score diverges. For our data, the upper limit according to this criterion is equal to $1/206.80=0.0048355$.
However, the dynamic win-lose score converges irrespective of the values of $\alpha$ and $\beta$ for the following reason. For expository purposes, let us assign different nodes to the same player at different times $t_n$ ($1\le n\le n_{\max}$). Then, \EQ\eqref{eq:def of W_{t_n}} implies that
any link in the network, which represents a game at time $t_n$,
is directed from the winner at $t_n$ to the loser at $t_{n}$ or earlier times.
Because there is no time-reversed link (i.e., from $t_n$ to $t_{n^{\prime}}$, where $t_n<t_{n^{\prime}}$) and any pair of players play at most once at any $t_n$, the network is acyclic.
The upper limit of $\alpha$
is infinite when the network is acyclic \cite{Park2005JSM}.
On the basis of this observation,
we examine the behavior of the dynamic win-lose score for
various values of $\alpha$.

In the official ATP ranking, the score
of a player is calculated from the player's performance in the last
52 weeks $\approx$ one year \cite{ATPWorldTour}. The results of the games in this time window
contribute to the current ranking of the player with the same weight if the other conditions are equal.
The dynamic win-lose score uses
the results of all the games in the past, and the contribution of the
game decays exponentially in time. By equating the contribution
of a single game in the two ranking systems, we assume $1\times 365 =
\int_{0}^{\infty } e^{-\beta t}dt$, which leads to $\beta = 1/365$.
In Results, we also investigated the robustness of the
ranking results against variations in
the $\alpha$ and $\beta$ values.

\section*{Acknowledgments}

We thank Mikio Ogawa for the assistance of data collection,
Juyong Park for valuable discussion, 
Taro Takaguchi for critical reading of the manuscript,
and the Association of Tennis Professionals for making the data set 
used in the present study publicly available. We acknowledge financial 
supports provided through Grants-in-Aid for Scientific Research (No. 23681033) and Innovative Areas ``Systems Molecular Ethology''(No. 20115009)) from MEXT, Japan. This research is also supported by the Aihara Project, the
FIRST program from JSPS, initiated by CSTP.


\begin{thebibliography}{10}
\expandafter\ifx\csname url\endcsname\relax
  \def\url#1{\texttt{#1}}\fi
\expandafter\ifx\csname urlprefix\endcsname\relax\def\urlprefix{URL }\fi
\providecommand{\bibinfo}[2]{#2}
\providecommand{\eprint}[2][]{\url{#2}}

\bibitem{Stefani1997JAS}
\bibinfo{author}{Stefani, R.~T.}
\newblock \bibinfo{title}{{Survey of the major world sports rating systems}}.
\newblock \emph{\bibinfo{journal}{J. Appl. Stat.}}
  \textbf{\bibinfo{volume}{24}}, \bibinfo{pages}{635--646}
  (\bibinfo{year}{1997}).

\bibitem{Daniels1969Biom}
\bibinfo{author}{Daniels, H.~E.}
\newblock \bibinfo{title}{Round-robin tournament scores}.
\newblock \emph{\bibinfo{journal}{Biometrika}} \textbf{\bibinfo{volume}{56}},
  \bibinfo{pages}{295--299} (\bibinfo{year}{1969}).

\bibitem{Moon1970SiamRev}
\bibinfo{author}{Moon, J.~W.} \& \bibinfo{author}{Pullman, N.~J.}
\newblock \bibinfo{title}{On generalized tournament matrices}.
\newblock \emph{\bibinfo{journal}{SIAM Rev.}} \textbf{\bibinfo{volume}{12}},
  \bibinfo{pages}{384--399} (\bibinfo{year}{1970}).

\bibitem{Borm2002AOR}
\bibinfo{author}{Borm, N.~E.}, \bibinfo{author}{Brink, R. V.~D.} \&
  \bibinfo{author}{Slikker, M.}
\newblock \bibinfo{title}{An iterative procedure for evaluating digraph
  competitions}.
\newblock \emph{\bibinfo{journal}{Ann. Operat. Res.}}
  \textbf{\bibinfo{volume}{109}}, \bibinfo{pages}{61--75}
  (\bibinfo{year}{2002}).

\bibitem{Saavedra2010PhysicaA}
\bibinfo{author}{Saavedra, S.}, \bibinfo{author}{Powers, S.},
  \bibinfo{author}{McCotter, T.}, \bibinfo{author}{Porter, M.~A.} \&
  \bibinfo{author}{Mucha, P.~J.}
\newblock \bibinfo{title}{{Mutually-antagonistic interactions in baseball
  networks}}.
\newblock \emph{\bibinfo{journal}{Physica A}} \textbf{\bibinfo{volume}{389}},
  \bibinfo{pages}{1131--1141} (\bibinfo{year}{2010}).

\bibitem{Radicchi2011PlosOne}
\bibinfo{author}{Radicchi, F.}
\newblock \bibinfo{title}{Who is the best player ever? {A} complex network
  analysis of the history of professional tennis}.
\newblock \emph{\bibinfo{journal}{PLoS ONE}} \textbf{\bibinfo{volume}{6}},
  \bibinfo{pages}{e17249} (\bibinfo{year}{2011}).

\bibitem{Callaghan2004NAMS}
\bibinfo{author}{Callaghan, T.}, \bibinfo{author}{Mucha, P.~J.} \&
  \bibinfo{author}{Porter, M.~A.}
\newblock \bibinfo{title}{{The bowl championship series: a mathematical
  review}}.
\newblock \emph{\bibinfo{journal}{Notices of the Am. Math. Soc.}}
  \textbf{\bibinfo{volume}{51}}, \bibinfo{pages}{887--893}
  (\bibinfo{year}{2004}).

\bibitem{Herings2005SCW}
\bibinfo{author}{Herings, P. J.~J.}, \bibinfo{author}{van~der Laan, G.} \&
  \bibinfo{author}{Talman, D.}
\newblock \bibinfo{title}{{The positional power of nodes in digraphs}}.
\newblock \emph{\bibinfo{journal}{Soc. Choice Welfare}}
  \textbf{\bibinfo{volume}{24}}, \bibinfo{pages}{439--454}
  (\bibinfo{year}{2005}).

\bibitem{Park2005JSM}
\bibinfo{author}{Park, J.} \& \bibinfo{author}{Newman, M. E.~J.}
\newblock \bibinfo{title}{{A network-based ranking system for US college
  football}}.
\newblock \emph{\bibinfo{journal}{J. Stat. Mech.}} \bibinfo{pages}{P10014}
  (\bibinfo{year}{2005}).

\bibitem{HolmeSaramaki2012PhysRep}
\bibinfo{author}{Holme, P.} \& \bibinfo{author}{Saram\"{a}ki, J.}
\newblock \bibinfo{title}{{Temporal networks}}.
\newblock \emph{\bibinfo{journal}{Phys. Rep.}} \textbf{\bibinfo{volume}{519}},
  \bibinfo{pages}{97--125} (\bibinfo{year}{2012}).

\bibitem{Tang2010SNS}
\bibinfo{author}{Tang, J.}, \bibinfo{author}{Musolesi, M.},
  \bibinfo{author}{Mascolo, C.}, \bibinfo{author}{Latora, V.} \&
  \bibinfo{author}{Nicosia, V.}
\newblock \bibinfo{title}{{Analysing information flows and key mediators
  through temporal centrality metrics}}.
\newblock In \emph{\bibinfo{booktitle}{Proceedings of the 3rd Workshop on
  Social Network Systems}} (\bibinfo{year}{2010}).

\bibitem{Pan2011PRE}
\bibinfo{author}{Pan, R.~K.} \& \bibinfo{author}{Saram\"{a}ki, J.}
\newblock \bibinfo{title}{{Path lengths, correlations, and centrality in
  temporal networks}}.
\newblock \emph{\bibinfo{journal}{Phys. Rev. E}} \textbf{\bibinfo{volume}{84}},
  \bibinfo{pages}{016105} (\bibinfo{year}{2011}).

\bibitem{Grindrod2011PRE}
\bibinfo{author}{Grindrod, P.}, \bibinfo{author}{Parsons, M.~C.},
  \bibinfo{author}{Higham, D.~J.} \& \bibinfo{author}{Estrada, E.}
\newblock \bibinfo{title}{{Communicability across evolving networks}}.
\newblock \emph{\bibinfo{journal}{Phys. Rev. E}} \textbf{\bibinfo{volume}{83}},
  \bibinfo{pages}{046120} (\bibinfo{year}{2011}).

\bibitem{Elo1978book}
\bibinfo{author}{Elo, A.~E.}
\newblock \emph{\bibinfo{title}{The Rating of Chess Players, Past \& Present}}
  (\bibinfo{publisher}{Arco}, \bibinfo{address}{New York},
  \bibinfo{year}{1978}).

\bibitem{Bradley1976Biom}
\bibinfo{author}{Bradley, R.~A.}
\newblock \bibinfo{title}{{Science, statistics, and paired comparisons}}.
\newblock \emph{\bibinfo{journal}{Biometrics}} \textbf{\bibinfo{volume}{32}},
  \bibinfo{pages}{213--232} (\bibinfo{year}{1976}).

\bibitem{Glickman1993thesis}
\bibinfo{author}{Glickman, M.~E.}
\newblock \bibinfo{title}{{Paired comparison models with time-varying
  parameters}}.
\newblock \emph{\bibinfo{journal}{PhD Dissertation. Department of Statistics,
  Harvard University, Cambridge}}  (\bibinfo{year}{1993}).

\bibitem{Fahrmeir1994JASA}
\bibinfo{author}{Fahrmeir, L.} \& \bibinfo{author}{Tutz, G.}
\newblock \bibinfo{title}{{Dynamic stochastic models for time-dependent ordered
  paired comparison systems}}.
\newblock \emph{\bibinfo{journal}{J. Amer. Stat. Asso.}}
  \textbf{\bibinfo{volume}{89}}, \bibinfo{pages}{1438--1449}
  (\bibinfo{year}{1994}).

\bibitem{Glickman1999JRSSSC}
\bibinfo{author}{Glickman, M.~E.}
\newblock \bibinfo{title}{{Parameter estimation in large dynamic paired
  comparison experiments}}.
\newblock \emph{\bibinfo{journal}{J. R. Stat. Soc. Ser. C}}
  \textbf{\bibinfo{volume}{48}}, \bibinfo{pages}{377--394}
  (\bibinfo{year}{1999}).

\bibitem{Knorrheld2000Stat}
\bibinfo{author}{Knorr-Held, L.}
\newblock \bibinfo{title}{{Dynamic rating of sports teams}}.
\newblock \emph{\bibinfo{journal}{J. R. Stat. Soc. Ser. D}}
  \textbf{\bibinfo{volume}{49}}, \bibinfo{pages}{261--276}
  (\bibinfo{year}{2000}).

\bibitem{Coulom2008LNCS}
\bibinfo{author}{Coulom, R.}
\newblock \bibinfo{title}{{Whole-history rating: a Bayesian rating system for
  players of time-varying strength}}.
\newblock \emph{\bibinfo{journal}{LNCS}} \textbf{\bibinfo{volume}{5131}},
  \bibinfo{pages}{113--124} (\bibinfo{year}{2008}).

\bibitem{Herbrich2007NIPS}
\bibinfo{author}{Herbrich, R.}, \bibinfo{author}{Minka, T.} \&
  \bibinfo{author}{Graepel, T.}
\newblock \bibinfo{title}{{TrueSkill${}^{\rm TM}$: a Bayesian skill rating
  system}}.
\newblock \emph{\bibinfo{journal}{Advances in Neural Information Processing
  Systems}} \textbf{\bibinfo{volume}{19}}, \bibinfo{pages}{569--576}
  (\bibinfo{year}{2007}).

\bibitem{Dixon1997ApplStat}
\bibinfo{author}{Dixon, M.~J.} \& \bibinfo{author}{Coles, S.~G.}
\newblock \bibinfo{title}{{Modelling association football scores and
  inefficiencies in the football betting market}}.
\newblock \emph{\bibinfo{journal}{J. R. Stat. Soc. Ser. C}}
  \textbf{\bibinfo{volume}{46}}, \bibinfo{pages}{265--280}
  (\bibinfo{year}{1997}).

\bibitem{ATPWorldTour}
\bibinfo{title}{{\rm http://www.atpworldtour.com} (date of access: October 1, 2011)}.

\bibitem{Martinich2002Inter}
\bibinfo{author}{Martinich, J.}
\newblock \bibinfo{title}{{College football rankings: do the computers know
  best?}}
\newblock \emph{\bibinfo{journal}{Interfaces}} \textbf{\bibinfo{volume}{32}},
  \bibinfo{pages}{85--94} (\bibinfo{year}{2002}).

\bibitem{Bennaim2006JQAS}
\bibinfo{author}{Ben-Naim, E.}, \bibinfo{author}{Vazquez, F.} \&
  \bibinfo{author}{Redner, S.}
\newblock \bibinfo{title}{{Parity and predictability of competitions}}.
\newblock \emph{\bibinfo{journal}{J. Quantitative Analysis in Sports}}
  \textbf{\bibinfo{volume}{2 (4)}}, \bibinfo{pages}{article 1}
  (\bibinfo{year}{2006}).

\bibitem{Lundh2006JQAS}
\bibinfo{author}{Lundh, T.}
\newblock \bibinfo{title}{{Which ball is the roundest? --- a suggested
  tournament stability index}}.
\newblock \emph{\bibinfo{journal}{J. Quantitative Analysis in Sports}}
  \textbf{\bibinfo{volume}{2 (3)}}, \bibinfo{pages}{article 1}
  (\bibinfo{year}{2006}).

\bibitem{Park2010JSM}
\bibinfo{author}{Park, J.}
\newblock \bibinfo{title}{{Diagrammatic perturbation methods in networks and
  sports ranking combinatorics}}.
\newblock \emph{\bibinfo{journal}{J. Stat. Mech.}} \bibinfo{pages}{P04006}
  (\bibinfo{year}{2010}).

\bibitem{Coleman2005Interface}
\bibinfo{author}{Coleman, B.~J.}
\newblock \bibinfo{title}{{Minimizing game score violations in college football
  rankings}}.
\newblock \emph{\bibinfo{journal}{Interfaces}} \textbf{\bibinfo{volume}{35}},
  \bibinfo{pages}{483--496} (\bibinfo{year}{2005}).

\bibitem{Brin98}
\bibinfo{author}{Brin, S.} \& \bibinfo{author}{Page, L.}
\newblock \bibinfo{title}{Anatomy of a large-scale hypertextual web search
  engine}.
\newblock \emph{\bibinfo{journal}{Proceedings of the Seventh International
  World Wide Web Conference}} \bibinfo{pages}{107--117} (\bibinfo{year}{1998}).

\bibitem{Kendall1938Biom}
\bibinfo{author}{Kendall, M.~G.}
\newblock \bibinfo{title}{{A new measure of rank correlation}}.
\newblock \emph{\bibinfo{journal}{Biometrika}} \textbf{\bibinfo{volume}{30}},
  \bibinfo{pages}{81--93} (\bibinfo{year}{1938}).

\bibitem{Fagin2003SIAMDM}
\bibinfo{author}{Fagin, R.}, \bibinfo{author}{Kumar, R.} \&
  \bibinfo{author}{Sivakumar, D.}
\newblock \bibinfo{title}{{Comparing top $k$ lists}}.
\newblock \emph{\bibinfo{journal}{SIAM J. Disc. Math.}}
  \textbf{\bibinfo{volume}{17}}, \bibinfo{pages}{134--160}
  (\bibinfo{year}{2003}).

\bibitem{Mccown2007JCDL}
\bibinfo{author}{McCown, F.} \& \bibinfo{author}{Nelson, M.~L.}
\newblock \bibinfo{title}{{Agreeing to disagree: search engines and their
  public interfaces}}.
\newblock In \emph{\bibinfo{booktitle}{Proceedings of the 7th ACM/IEEE-CS Joint
  Conference on Digital Libraries}}, \bibinfo{pages}{309--318}
  (\bibinfo{year}{2007}).

\end{thebibliography}

\section*{Author contributions}

N.M. designed the research; S.M. contributed the computational results; S.M. and N.M. discussed the results; N.M. wrote the paper.

\section*{Additional information}

\textbf{Competing financial interests:} The author declares no competing financial interests.

\newpage
\clearpage

\clearpage

\clearpage

\clearpage

\clearpage

\clearpage

\end{document}